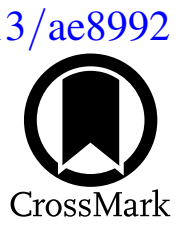

# New Statistical Topology Theory Predicts Turbulent Magnetic Emergence from the Sun's Interior

Anda Xiong[1,2], Hongyan Li[1,2], Shangbin Yang[1,2], Haiqing Xu[1,2], Quan Wang[1,2], Haisheng Ji[3], Xin Liu[4], Yuanyong Deng[1,2], and Hongqi Zhang[1,2]

[1] State Key Laboratory of Solar Activity and Space Weather, National Astronomical Observatories, Chinese Academy of Sciences, Beijing 100101, People's Republic of China; adxiong@nao.cas.cn, yangshb@nao.cas.cn
[2] University of Chinese Academy of Sciences, Beijing 100049, People's Republic of China
[3] Purple Mountain Observatory, Chinese Academy of Sciences, Nanjing 210023, People's Republic of China
[4] School of Physics and Optoelectronic Engineering, Beijing University of Technology, Beijing 100124, People's Republic of China

## Abstract

We propose and verify a new statistical topology framework to study the complex magnetic field evolution of Sun-like stars. The Sun, as the star we are most familiar with, exhibits chaotic behaviors such as solar flares and mass ejections that are crucial to the Earth. While these phenomena are mainly driven by the magnetic field, it has been challenging to understand the complex magnetic field. In this paper, we propose a new model to understand the helicity behavior of magnetic loops before their emergence from the interior by advancing the loop ensemble theory from statistical physics. We derive several new power-law scalings that are essential to the Sun's magnetic field, including magnetic flux, magnetic helicity, and linking number. We examine our prediction by a large data analysis through long-term continuous observation over 32 yr. These results not only provide evidence for the new statistical topology framework but also systematically explain the intrinsic unpredictability on the emergence of extreme solar activities. This new discovery on the critical structure of loop ensemble can also be applied to a wide range of turbulence systems.

## 1. Introduction

Among vast plasma systems from fusion device to astrophysics, the Sun stands out as a natural laboratory enabling observations of complex magnetohydrodynamics of proper spatial and temporal scale. The Sun's magnetic activity—governing outstanding phenomena such as solar flares, coronal mass ejections, and space weather—is fundamentally based on the dynamics of magnetic flux tubes. These twisted bundles of frozen magnetic field lines, emerging from the internal flow field, serve as the building blocks of active regions and the corona. While the structure and evolution of magnetic flux tubes of individual active region have been extensively studied, there is still a lack of global statistical description of them, which is particularly important for understanding a complex system like the Sun. Furthermore, the generation of solar magnetic field has long been a challenging question, whereas the dynamo theory points out the crucial role of knotted flow in the internal flow field, such that the magnetic flux tubes are stretched and twisted by the differential rotation and convective turbulence, known as the $\Omega$ and $\alpha$ effect. Since we are unable to directly observe this multiscale complex process in the solar internal flow field, one way to speculate its structure and dynamics, apart from helioseismology, could be deductions from the emerging magnetic activities. Therefore, a systematic framework to understand the statistical behavior of emerging magnetic activities can also allow one to infer the physics in the internal flow field, where the magnetic loops are formed before emergence.

To understand such a complex magnetic field, it is often considered by analyzing the tangled magnetic loops, because due to the divergence-free nature of the magnetic field, the field lines are topologically equivalent to loops. A critical measurement of such complex tangled topology is provided by magnetic helicity (M. Berger & G. Field 1984; M. Berger 1999), which has been shown as an ideal invariant (L. Woltjer 1958; K. Moffatt 1969) that quantifies the degree of knottiness within the field volume. While local knotted structures and helicity have been identified (M. Dennis et al. 2010; D. Kleckner & W. T. M. Irvine 2013) and measured (S. Yang & J. Buchner 2013; M. W. Scheeler et al. 2017) in various systems, a statistical description to the collective behavior of these tangled vortices remains elusive due to the lack of theoretical frameworks. The statistical topology of field lines is particularly crucial for understanding turbulence in multiscale systems, where knotted structures emerge and evolve across a broad range of spatial and temporal scales, such as in the Sun-like stars.

We establish a statistical topology theory for the complex magnetic field through an advanced loop soup framework (T. Vachaspati & A. Vilenkin 1984; G. F. Lawler & W. Werner 2003; K. O'Holleran et al. 2008; K. O'Holleran et al. 2009; A. Nahum & J. T. Chalker 2011; S. Sheffield & W. Werner 2012; A. Nahum et al. 2013; A. J. Taylor & M. R. Dennis 2016; A. J. Taylor 2018; A. Xiong et al. 2025) and find new asymptotic scaling laws for complex structures in turbulence. The loop soup model consists of a statistical ensemble of loops and has been studied in a broad range of systems such as quantum chaos (A. J. Taylor & M. R. Dennis 2016; A. J. Taylor 2018), optics (K. O'Holleran et al. 2008, 2009), and cosmology (T. Vachaspati & A. Vilenkin 1984). There have been mathematical analysis (G. F. Lawler & W. Werner 2003; S. Sheffield & W. Werner 2012) and rigorous derivations on the lattice model in condensed matter physics (A. Nahum & J. T. Chalker 2011; A. Nahum et al. 2013). Most previous results focused on the loop length distribution (T. Vachaspati & A. Vilenkin 1984; K. O'Holleran et al. 2008; A. Nahum et al. 2013) and found such distribution to be universal among these different physical systems. In this paper, we discover new power-law scalings for the magnetic topology, including the probability distribution of magnetic flux, linking number, and magnetic helicity, which are essential physical quantities to the solar magnetic activities and the knottiness in the dynamo process. We verify our theory by a long-term large data observation over 32 yr that continuously records the vector magnetograms of active regions on the Sun. Utilizing both theoretical framework and data quality far exceeding our previous calculation (A. Xiong et al. 2025), we for the first time characterize the critical behaviors of solar activities with these new discoveries. We find that the magnetic flux probability distribution follows a power law with convergent expectation value, suggesting that the magnetic flux in the corona might be governed by turbulence of multiple spatial scales. On the other hand, we find that the power-law distribution of magnetic helicity does not have a convergent expectation value, which means that the total knottiness of the solar magnetic field is significantly influenced by rare but extreme events. This new framework and result shows the intrinsic sensitivity of global magnetic topology and provides a basis to study and predict the global statistical behavior of solar magnetic activities.

## 2. Helicity and Loop Ensemble in the Sun

We exploit a long-term continuous observation data from Huairou Solar Observing Station (HSOS) of National Astronomical Observatories, Chinese Academy of Sciences, over 32 yr that covers three solar cycles. The observation is carried on the solar magnetic field telescope (G. X. Ai & Y. F. Hu 1986), which is a 35 cm aperture narrowband filter magnetograph on Fe I $\lambda = 5324$ Å for photospheric vector magnetic field and line-of-sight velocity field and H$\beta \lambda = 4861$ Å for chromospheric line-of-sight magnetic field and line-of-sight velocity field data. Each pixel on the telescope covers about $0\farcs38$ on the Sun (source image resolution), and the field of view is about $6' \times 4'$ days. We consider solar magnetogram data from 1988 to 2019 of 3248 active regions with 58,942 samples of magnetograms. For the best data quality, we only choose active regions within 30° of central meridian and represent each active region by a magnetogram with the best signal-to-noise ratio. Thus we in total obtained a number of 1490 magnetograms for active regions. We process these data by subtracting the polarized background and resolve the 180° ambiguity in the transverse magnetic field by using the potential field extrapolation method (acute angle method). We show in in the left panel of Figure 1 a sample vector magnetogram from observation and in the right panel the latitude distribution of these active regions. On both the north and south hemisphere, the latitude distribution appears as a slightly skewed Gaussian distribution. While the skewness comes from the differential rotation, the Gaussian distribution of the latitude suggests that the magnetic emergence follows a Gaussian process in total time accumulation. In solar physics, the famous "butterfly diagram" depicts the migration of emergence latitude from high to low, such that in the beginning of the solar cycle the active regions are more likely to appear in high latitude, while during the evolution they become to emerge in lower latitude. Meanwhile if we consider the total events accumulated in time, we find a Gaussian distribution on the latitude for magnetic emergence. The long-term observation to active regions allows analysis to the statistical behavior of the solar magnetic field's complex evolution.

To quantify such complex magnetic topology, we study the probability distribution of magnetic flux $\phi$, the linking number $n$ that characterizes the degree of knottiness of the flux tube, and the magnetic helicity $H_M$, which is an ideal invariant (L. Woltjer 1958) and essential to solar activities (M. Berger & G. Field 1984; M. Berger 1999). Helicity measures the degree of knottiness of the field in a broad range of systems, such as the kinetic helicity $H_k = \int_V \boldsymbol{u} \cdot (\nabla \times \boldsymbol{u})$ for fluids (K. Moffatt 1969), where $u$ is the velocity and the magnetic helicity

$$H_M = \int_V \boldsymbol{A} \cdot \boldsymbol{B} dV \tag{1}$$

for $\boldsymbol{B}$ is the magnetic field and $\boldsymbol{A}$ the potential. It has been shown that helicity is also the only integral invariant for volume-preserving transformation (A. Enciso et al. 2016) and equals the abelian Chern–Simons action (E. Witten 1989). Since $\boldsymbol{A}$ depends on the choice of gauge, in practice it is more convenient to consider the relative helicity (J. M. Finin & T. M. Antonsen 1985)

$$H_R = \int_V (\boldsymbol{A} + \boldsymbol{A_P}) \cdot (\boldsymbol{B} - \boldsymbol{P}) dV, \tag{2}$$

where $\boldsymbol{P}$ is the reference potential field and $\nabla \times \boldsymbol{A_P} = \boldsymbol{P}$. This expression finds the relative helicity compared between the active region and the background magnetic field. Since the background magnetic field is much smaller and negligible, the relative helicity can be seen approximately as the net magnetic helicity of the active region. We apply the nonlinear force-free field (NLFFF) method (T. Wiegelmann 2004; T. Wiegelmann et al. 2017) to extrapolate the three-dimensional magnetic field from the observed magnetogram. Since the real observation data of photospheric magnetic field are not perfectly force free (X. M. Zhang et al. 2017), the data are preprocessed according to T. Wiegelmann et al. (2006) and C. J. Schrijver et al. (2006). We also apply the multigrid method (D. MacTaggart et al. 2013; S. A. Gilchrist et al. 2016; X. Zhu & T. Wiegelmann 2022) to enhance the calculation performance by increasing the speed and stability. After obtaining the three-dimensional magnetic field data from NLFFF extrapolation method based on observed magnetograms, we calculate the magnetic helicity by applying the Coulomb–Yang method (S. Yang & J. Buchner 2013; S. Yang et al. 2018).

The helicity of knotted tubes can also be equivalently expressed by (K. Moffatt & R. Ricca 1992)

$$H = n\phi^2, \tag{3}$$

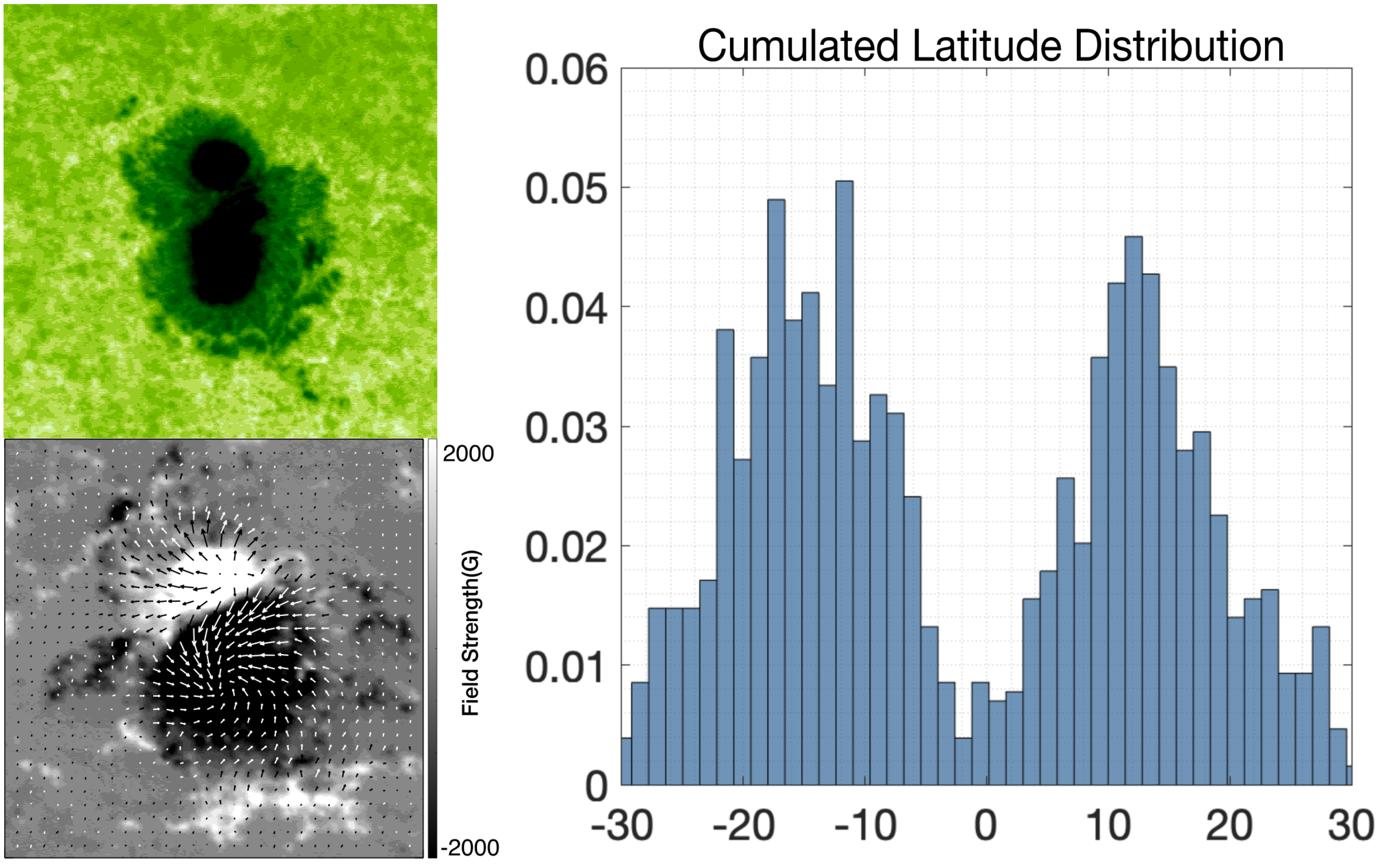

**Figure 1.** Left: a sample vector magnetogram from HSOS (AR 6659, in the year 1991). The green color indicates the wavelength 5324 Å and shows a clear bipolar structure. The grayscale image depicts the magnetic field strength in gausses, with the black and white arrows marking the direction of the transverse magnetic field. Right: the latitude distribution of our total 1490 sample active regions, which appears as a slightly skewed normal distribution.

where $n$ is the linking number and $\phi$ the flux. In general, the linking number $n$ of a system is the sum of self-linking number of each knotted structure and the interlinking number between arbitrary two structure components. Simple bipolar active regions are commonly seen in equatorial zone, if we approximate the bulk as a tube, then the linking number in this active region is mainly contributed by the self-linking number of the magnetic flux tube. The self-linking number equals the sum of twist and writhe, as found by the Călugăreanu's theorem (also known as the White's theorem; K. Moffatt & R. Ricca 1992). Since the self-linking number of certain knotted structure comes from the twist and writhe, such theorem is applied to knotted structures in a general form, including both knots and links. For magnetic flux tubes, since they are wedged to two poles, topologically they are equivalent to loops and knots, and the knottiness can be found equivalently between these knotted structures (M. Berger 1999). Therefore, the linking number in Equation (3) has been acknowledged as a critical measurement to magnetic helicity and provides a basis to further studies on magnetic helicity and flux such as M. Berger (1999), B. J. LaBonte et al. (2007), H. Jeong & J. Chae (2007), and S. Yang et al. (2009).

The magnetic helicity $H_M$ measures the degree of knottiness of the flux tube, and it has been shown that the helicity can affect the magnetic activity by inducing eruptions with the increase of the topological instability (M. Berger 1999; R. L. Moore et al. 2001; T. Török & B. Kliem 2005). In Figure 2 we present a sample of magnetic loops in the active region with high linking number $n$, meaning that in this region the magnetic field is strongly knotted. In active regions, the magnetic helicity is mainly contributed by the magnetic flux, and the linking number is usually less than 1, because when the linking number is high, the magnetic ropes are more likely to erupt. The magnetic loops are first formed in the internal flow field and then emerge into the atmosphere. Consequently, the knottiness of active region magnetic field is also related to global magnetic activity and dynamo pattern (H. Zhang et al. 2010a, 2010b), where the magnetic chirality has been shown as a reflection of the dynamo process. Since the convective zone is nearly well mixed, the magnetic loops in global distribution can be seen together as a statistical ensemble, but they are not observable until later emerging as active regions and provide evidence for the theory.

To study this complex statistical behavior of solar magnetic topology, we set up a new framework of statistical topology analysis. Compared to traditional statistical physics that studies the the position and momentum of particles, statistical topology focuses on topological structures, such as loops, as the fundamental object for analysis and studies the physical quantities that distribute with these structures. It is a divergence-free nature of magnetic field that the field lines are closed loops; thus with the scale-invariant nature of turbulence, the complex magnetic field in the internal flow field can be seen as consisting of loops in different scales, which could be referred to as a loop ensemble or loop soup (T. Vachaspati & A. Vilenkin 1984; K. O'Holleran et al. 2008;

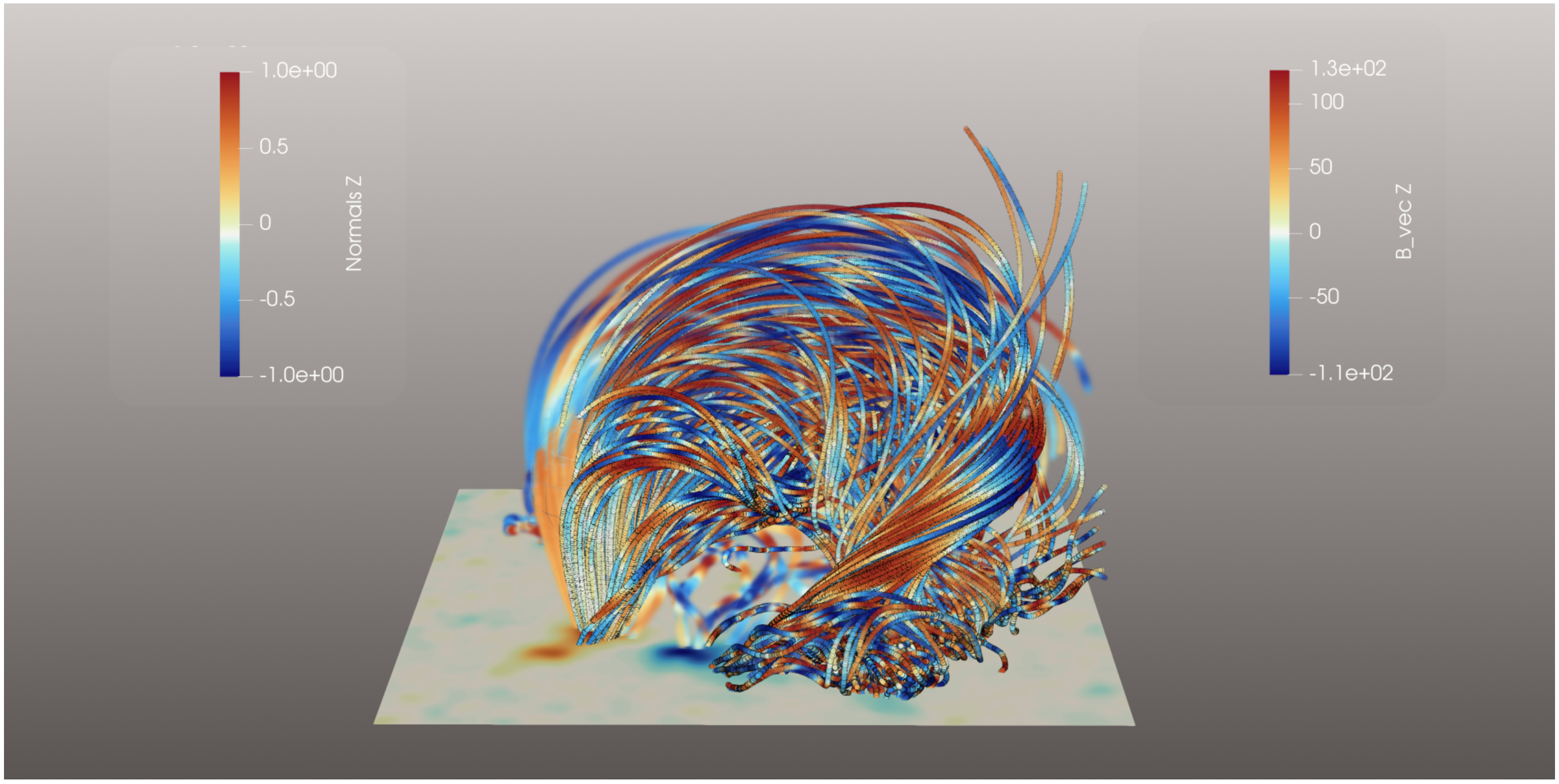

**Figure 2.** A sample active region with measured linking number $n \approx 1.4357$, which is much larger than the linking number of other usual active regions. This region labels NOAA 12697. This figure is plotted in ParaView based on three-dimensional magnetic field extrapolation.

A. Nahum & J. T. Chalker 2011; S. Sheffield & W. Werner 2012). This statistical perspective of singular structures such as vortices and field lines can be found from a broad range such as turbulence, quantum turbulence, and quantum chaos. While the governing equation varies among systems, and it is certainly true that the vortices may have different geometries and dynamics in different systems, they nonetheless can be seen from this perspective as a statistical ensemble of topological structures. We consider the magnetic structures in the internal flow field as an ensemble of magnetic loops, and when they emerge as active regions they become observed, and this process mathematically serves as a consistent sampling to the internal flow field. Therefore, the statistical distribution of active region magnetic topology suggests the loop soup statistics in the internal flow field.

The theoretical analysis starts from the principle of scale invariance. Today the idea of fractal and scale invariance has become familiarized by physical sciences, where a similar pattern or action persists through different length scales. By denoting $N$ as the number density and $V_\xi$ the characteristic volume, the principle of scale invariance reads

$$N \cdot V_\xi = N \cdot \xi^d = \text{const.}, \tag{4}$$

where $\xi$ is the characteristic length and $d$ is the dimension. In a loop soup like the magnetic loops in solar convection turbulence, the loops are formed in scale invariance, which means that the loops in different scales are formed under a similar mechanism, so naturally they build up a fractal set of loops. For loops in three-dimensional space, $N \propto \frac{1}{V_\xi} \propto \frac{1}{\xi^3}$, and thus $dN \propto \xi^{-4} d\xi$.

In this magnetic loop ensemble, we first find the power-law probability distribution of magnetic flux $\phi$ that measures the strength of magnetic field in a given region. By definition, $\phi = \int \boldsymbol{B} ds$, where $s$ is the cross-section area of the flux tube. To derive power-law scaling, it is effective to apply dimensional analysis. From dimensional analysis, the cross-section area of a tube is proportional to the square of length, $[s] = \xi^2$. So for an averaged magnetic field, we can approximately find $[\phi] \propto \xi^2$. Thus together with the principle of scale invariance in Equation (4), we find the probability distribution function of magnetic flux as

$$dN = C_\phi \phi^{-5/2} d\phi. \tag{5}$$

This result shows the asymptotic power-law scaling distribution of magnetic flux carried by loops. This power-law distribution of magnetic flux is derived solely based on the principle of scale invariance and dimensional analysis, without further assumption on the turbulence spectrum or vortex geometry. The statistical distribution of magnetic flux does not violate the principle of flux conservation because in the loop ensemble, the flux is distributed with the formation of loops in different length scales, while the conservation of flux applies for a single tube whose flux does not vary with continuous deformation. When the topology of the tube is local open lines instead of closed loops, it has a different statistical distribution (A. Nahum & J. T. Chalker 2011; A. Nahum et al. 2013; A. J. Taylor & M. R. Dennis 2016; A. J. Taylor 2018). Since the open field lines are locally extending in one-dimension, the principle of scale invariance gives $N \propto 1/\xi$, and with $[\phi] \propto \xi^2$ from dimensional analysis, one finds $dN = C_\phi \phi^{-3/2} d\phi$. Therefore, there is no conflict between our result and previous statistics on the power-law distribution of magnetic flux (C. E. Parnell et al. 2009), which gives the value for around $-1.85 \pm 0.14$, which is in between our theoretical value of $-5/2$ for loops and $-3/2$ for local open lines, whereas the magnetic topology type in C. E. Parnell et al. (2009) was not specified. Furthermore, sometimes the magnetic topology of active regions can be even more complex, such as being composed by multiple loops wedged to different positive and negative poles, and that would also lead to a different distribution.

We then find the probability distribution of the linking number $n$ by studying the local geometry in this loop soup. It has been shown that the structures in turbulence can be generally described by Schramm–Loewner evolution curves (D. Bernard et al. 2006; S. Thalabard et al. 2011), which is a family of random fractal curves. This parameterization is neither against the deterministic governing equation in the flow field nor ignoring the correlations, but it is a mathematical framework to describe the geometry of the curves. It is known that the internal flow field is highly complex and turbulent and cannot be modeled by a single conventional turbulence model. In a review (Y. Fan 2021), it has been realized that the rise and emergence of magnetic structures from the convection zone is a complex interplay between global toroidal field and local turbulence. It is also known that the internal flow field turbulence cannot be modeled by a single conventional turbulence model. First, due to the buoyancy, the turbulence is no longer isotropic, such that beside the Kolmogorov's spectrum, there is the Bolgiano–Obukhov spectrum (A. Brandenburg 1992) in the convection, and this makes the internal flow field a hybrid of turbulences. Furthermore, the solar convection differs from typical Rayleigh–Benard convection by having an open boundary, and consequently the convection cells are turbulent and irregular. Given these conditions, it is convenient to apply Brownian geometry as an estimate to the local geometry of this loop soup. It is not misinterpreting turbulence as some kind of noise but only an effective method for local geometry estimation. Then the geometry of Brownian curve gives $\xi \sim \sqrt{l}$, where $l$ is the arc length of certain curve segment. Then with a simple calculation one finds the relation between $N$ and $l$ as $dN \propto l^{-5/2}dl$ for the loop length distribution in Brownian loop soup (T. Vachaspati & A. Vilenkin 1984; K. O'Holleran et al. 2008). This shows the asymptotic scaling of the loop length distribution, such that when the length of a loop increases, the probability to find it decreases in an asymptotic power law, such that there are many small loops and less large loops. It is worth pointing out that besides loops, the field lines or vortices may also have the topology of local open curves that virtually close at infinity, which in periodical boundary condition would become very long lines penetrating through the local space, and appear as the "longest loops," sometimes referred to as "nontrivial homology lines" (A. J. Taylor & M. R. Dennis 2016; A. J. Taylor 2018). While the loop topology appears as "coronal loops" in the Sun, these local open curves appears as "coronal holes," which are more likely to be observed in high-latitude regions. For local open field lines, the length distributions have a different power-law scaling (A. Nahum & J. T. Chalker 2011; A. Nahum et al. 2013; A. J. Taylor & M. R. Dennis 2016; A. J. Taylor 2018) because they are so long that they can be approximately seen as extending in one-dimension ($d = 1$), which leads to $dN \propto l^{-3/2}dl$. From the loop length distribution, it is simple to find the probability distribution of the linking number, since it has been proved (E. Panagiotou et al. 2009) that statistically the average linking number increases linearly with the random loop length; therefore we would expect

$$dN = C_n n^{-5/2} dn \tag{6}$$

for the asymptotic scaling behavior for the linking number $n$ in a random loop ensemble. This result suggests that the probability to find a certain value of the linking number of a random loop decreases in an asymptotic power law with index $-5/2$. The knot probability of random loops has been more extensively studied in T. Deguchi & K. Tsurusaki (1997) and A. Xiong et al. (2021, 2025), where a universal knot probability equation for closed random loops was found, but here for the concern of helicity it is sufficient to focus on the linking number primarily.

Based on the probability distribution of $\phi$ and $n$, we find the probability distribution of helicity $H$ carried by the loops. From Equation (3) is directly written

$$\frac{dN}{dH} = \frac{dN}{dn}\frac{\partial n}{\partial H} + \frac{dN}{d\phi}\frac{\partial \phi}{\partial H}. \tag{7}$$

While both $n$ and $\phi$ have similar power-law distribution, we find that the magnitude of $n$ is much smaller than $\phi$, which is consistent with previous estimate on the magnitude of $n$ (B. J. LaBonte et al. 2007; S. Yang et al. 2009). Then the calculation can be simplified by considering the helicity being mainly contributed by flux ($n \ll \phi$), thus

$$\begin{aligned}\frac{dN}{dH} &\approx \frac{dN}{d\phi}\frac{\partial \phi}{\partial H}\\ &\approx C_H H^{-7/4}.\end{aligned} \tag{8}$$

In Figure 3 we verify these theoretical predictions by a long continuous observation data on the last three solar cycles, with a number of 1490 observed active regions of high-quality magnetograms. In the main window we plot the asymptotic power-law scaling of the magnetic helicity distribution in double natural log scale, and in the inset window we plot the distribution of linking number $n$ and magnetic flux $\phi$. The distribution of magnetic helicity lasts for nearly 4 magnitudes in natural log scale, and the data stop around $e^{100} \approx 2.68 \times 10^{43}\mathrm{Mx}^2$, which is near the upper bound for typical active regions, since when the magnetic helicity grows larger, the flux ropes become more easy to erupt. There are extreme events where the magnetic helicity can reach up to $10^{44}\mathrm{Mx}^2$, but they do not appear on the plot since they are statistically rare. The magnetic flux of each magnetogram is calculated as the mean from each pixels' flux. Magnetic flux of magnetograms are generally in magnitude of $10^{21}$–$10^{23}$ Mx, so in the measurement there is not a significant effect from the noise. We find $n$ has its power-law distribution for a small value of less than 1 as when the linking number is high, the strong twist and writhe may commonly lead to topology change and eruption. In this scaling region, the linking number ranges over 2 magnitudes in a natural logarithm, which is nearly 2 magnitudes in a common logarithm. According to E. Panagiotou et al. (2009), in the characteristic scale, linking number scales linearly proportional to the loop size, suggesting that the active region size should also span over at least 2 magnitudes in a common logarithm. This agrees with previous statistics (K. L. Harvey & C. Zwaan 1993) on the scaling range of active region size. However, so far the expansion of magnetic loops during emergence due to the turbulence is not yet comprehensively clear, and it is possible that the loops do not all expand in the same ratio. Therefore, it is possible for an active region to have small size but large linking number as the active region size is not necessarily proportional to the magnetic loop size in the internal flow field. While the statistical distribution of active region magnetic topology is verified, the individual evolution and expansion remain unclear and are worth further study in future.

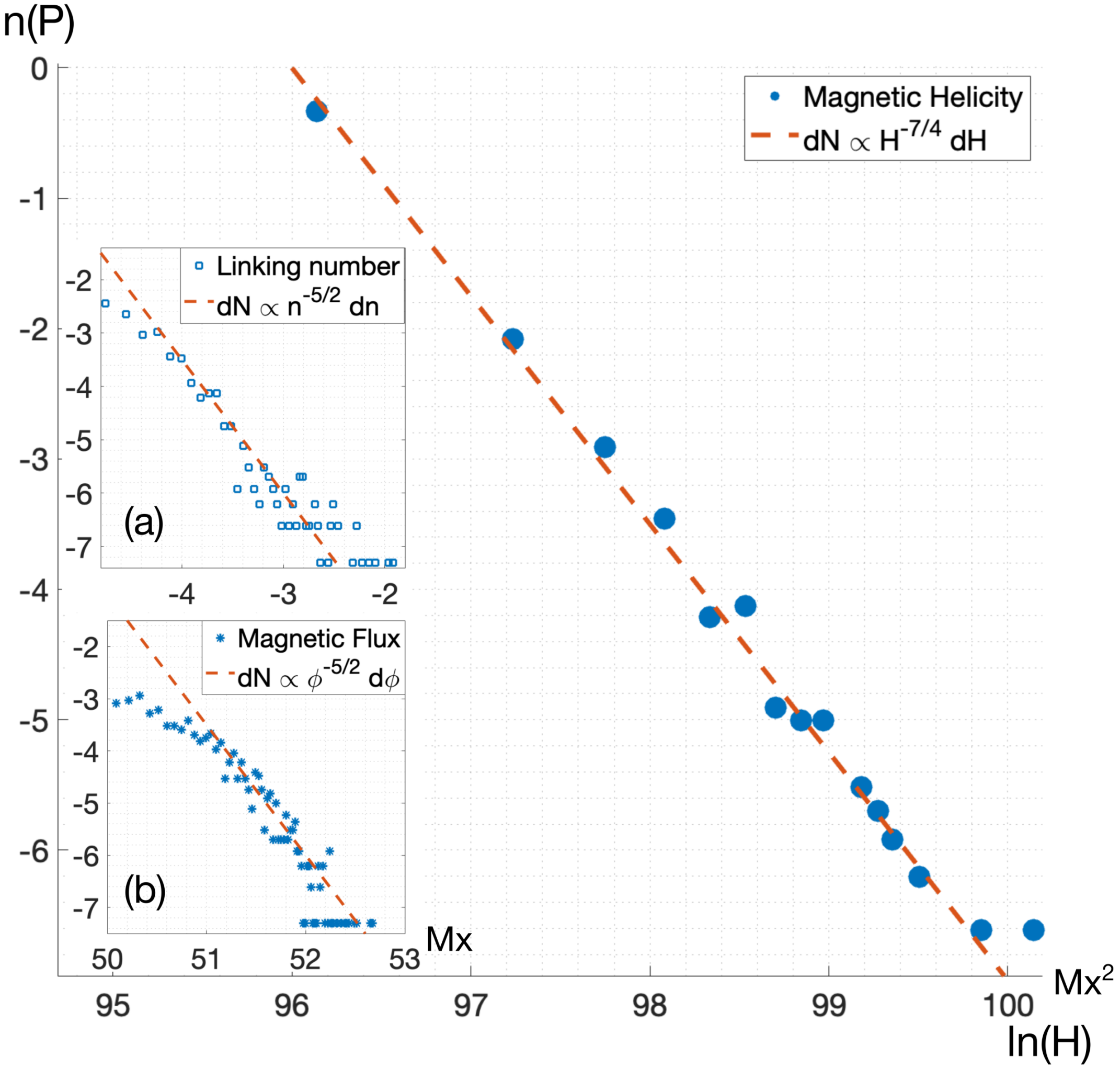

**Figure 3.** Asymptotic power-law distribution of helicity $H$ (in the main window), linking number $n$ (in subpanel (a)), and magnetic flux (in subpanel (b)). The histogram of the data is plotted in double natural log scale, and it is evident that the slope agrees with our theoretical prediction. In the plot, magnetic helicity has a scaling range from exp(96) to exp(100), which is from $10^{41}$ to $10^{43}$Mx$^2$. The magnetic flux ranges from around exp(51) to exp(53) Mx, which is nearly from $10^{21}$ to $10^{23}$ Mx. The linking number ranges from exp(−4.5) to exp(−2), which is from around $10^{-3}$–$10^{-1}$. In physics, the scaling range could extend beyond the plot, but it would then require a greater amount of data for statistics.

The power-law scaling of magnetic helicity is for the first time found by the new statistical topology theory. It is also important to acknowledge the pioneering studies on the power-law scalings from previous statistical theories such as self-organized criticality (M. J. Aschwanden & C. E. Parnell 2002; D. Hughes & M. Paczuski 2003; M. J. Aschwanden 2022; M. J. Aschwanden & N. V. Nhalil 2023), percolation (D. G. Wentzel & P. E. Seiden 1992), random walk (H. J. Hagenaar et al. 1999), and turbulence (V. I. Abramenko et al. 2012), which are fundamental principles and universal statistical models for complex systems. In order to better understand the complex magnetic structure of active regions, it is helpful to specify its topology. Previous studies derived the power-law distribution index for magnetic flux to be –1.7 (D. Hughes & M. Paczuski 2003) or –1.8 (M. J. Aschwanden & N. V. Nhalil 2023) for general magnetic emergence under self-organized criticality, while in this paper we specify the topological structure and find such index to be –2.5 for loops. From Equation (8) it is direct that the $-5/2$ index for magnetic flux distribution and the $-7/4$ index for magnetic helicity distribution are sufficient and necessary conditions to each other; therefore the new scaling law for helicity distribution verifies the new statistical topology theory as a whole, which is supported by the data.

## 3. Critical Behavior of Magnetic Emergence

Based on the discovery of these probability distributions, we discuss their implications on the structure and evolution of solar activities. Generally, for a given variable $x$ in probability distribution function $P(x) = Cx^{-\alpha}$, the expectation value $\langle x \rangle$ is

$$\langle x \rangle = \int_{x_{\min}}^{x_{\max}} xp(x)dx = \frac{C}{2-\alpha} x^{2-\alpha}\Big|_{x_{\min}}^{x_{\max}}. \quad (9)$$

Therefore, the value of $\alpha$ determines the statistical type of distribution, since the expectation value converges only when $\alpha$ is greater than 2. When $\alpha$ is greater than 2, the distribution has a convergent expectation value and is dominated by small-scale events; on the other hand, when $\alpha$ is less than 2, the distribution does not have a convergent expectation value and is influenced by extreme events with small probability but large impact. For magnetic flux distribution, $\alpha = 5/2$, which is greater than 2, suggesting that the emergence of magnetic flux has a finite expectation value and is dominated by small-scale events. The emergence of flux ropes from convection zone is a result of magnetic buoyancy, and our result suggests that the statistics of this process is predictable and dominated by the behavior of flux tubes with a small amount but large number. On the other hand, for magnetic helicity the power-law index $\alpha = 7/4$ is less than 2, suggesting that the mean value of magnetic helicity in active regions does not converge to a finite bound. Thus the global magnetic helicity distribution is statistically significantly influenced by extreme events. Extreme events with a very large amount of helicity can be the main contributor to the global total helicity and can influence the total solar magnetic topology. Since the magnetic topology leads to solar eruptions, the power-law scaling index of the magnetic helicity distribution explains the intrinsic unpredictability on its average value.

In this new statistical topology framework, we consider the internal flow field as an ensemble of magnetic loops and derive the power-law distribution of flux and helicity solely based on basic principles of scale invariance and dimensional analysis. Considering the complex interplay between global toroidal field and local turbulence, we then further apply a Brownian estimation to the local geometries of the loops and find the distribution of linking number. Then as the magnetic structures emerge, the observations to them is mathematically a consistent sampling to examine such predicted distribution. We verify these predictions by a large sample of data that cover three solar cycles over 32 yr. We also point out that these distribution laws are universal and are not to be limited within solar physics, and we expect them to be verified in a broad range of systems such as fluid, superfluid, plasmas, and many other systems if they can be modeled by the loop soup ensemble. We for the first time link these power laws to the structure of the Sun and point out that the global magnetic topology distribution can be influenced by a single extreme event.

## Acknowledgments

We thank Renzo L. Ricca, Mark R. Dennis, Rodion Stepanov and Xing Wei for the valuable discussions. We also thank the reviewers for the valuable comments and suggestions.

This research is supported by the National Key R&D Program of China Nos. 2022YFF0503800, 2021YFA1600500, and 2022YFF0503001; National Natural Science Foundation of China (grant Nos. 11427901, 12250005, 12073040, 12273059, 11973056, 12003051, 11573037, 12073041, 11427901, 11572005, 12250014, and 11611530679); the Strategic Priority Research Program of the Chinese Academy of Sciences (grant Nos. XDB0560000, XDA15052200, XDB09040200, XDA15010700, XDB0560301, and XDA15320102); Beijing Natural Science Foundation (grant Nos. IS23030 and Z180007), China's Space Origins Exploration Program and the Chinese Meridian Project (CMP).

## Author Contributions

Anda Xiong conceived and wrote the paper. Hongyan Li contributed as co-first author. Hongyan Li, Shangbin Yang, Haiqing Xu, Quan Wang, Yuanyong Deng, and Hongqi Zhang performed observations and developed the calculation method. Haisheng Ji and Xin Liu contributed to the formulation of the paper.

## ORCID iDs

Anda Xiong https://orcid.org/0009-0002-6844-2801
Shangbin Yang https://orcid.org/0000-0002-2967-4522
Haiqing Xu https://orcid.org/0000-0003-4244-1077
Quan Wang https://orcid.org/0000-0003-3142-217X
Haisheng Ji https://orcid.org/0000-0002-5898-2284
Yuanyong Deng https://orcid.org/0000-0003-1988-4574